\documentclass[10pt,conference]{IEEEtran}
\IEEEoverridecommandlockouts
\usepackage[noadjust]{cite}
\usepackage{amsmath,amssymb,amsfonts}
\usepackage{algorithmic}
\usepackage[dvipdfmx]{graphicx}
\usepackage{textcomp}
\usepackage{xcolor}
\usepackage{url}
\usepackage{bm}
\def\BibTeX{{\rm B\kern-.05em{\sc i\kern-.025em b}\kern-.08em
    T\kern-.1667em\lower.7ex\hbox{E}\kern-.125emX}}
\begin{document}

\title{Self-Organizing Score-based Data Assimilation}

 \author{\IEEEauthorblockN{ Yuma Yamaoka, ~~~Seiichi Uchida, ~~~Shoji Toyota}
 \IEEEauthorblockA{\textit{Department of Advanced Information Technology}, 
 \textit{Kyushu University},  \\
 Fukuoka, Japan \\
 {\{yuma.yamaoka@human., uchida@, toyota@\}ait.kyushu-u.ac.jp}}
}

\maketitle

\begin{abstract}
A state-space model is a statistical framework for inferring latent states from observed time-series data. However, inference with nonlinear and high-dimensional state-space models remains challenging. To this end, an approach based on diffusion models—a powerful class of deep generative models—has been developed, known as {\em Score-based Data Assimilation} (SDA)~\cite{NEURIPS2023_SDA}. However, SDA cannot be directly applied when the latent-state transition depends on unknown parameters that must be inferred jointly with the latent states. To overcome this limitation, we propose a framework that enables SDA to handle latent states with unknown parameters.
A key feature of the proposed method is the incorporation of the self-organization technique~\cite{Self_Organizing}, which has been used in classical state-space modeling for the joint estimation of latent states and parameters.
By integrating this classical technique into modern SDA, our method enables joint inference of latent states and unknown parameters while maintaining the high training efficiency of SDA. 
The effectiveness of the proposed approach is validated through numerical experiments on dynamical systems arising in neuroscience and atmospheric science. In addition, its scalability is demonstrated using a high-dimensional Kolmogorov flow, with the data dimension on the order of several hundred thousand.

\end{abstract}


\section{Introduction}

Given a time series of observations $\bm{y}_1, \ldots, \bm{y}_T$, recovering the underlying latent states $\bm{x}_1, \ldots, \bm{x}_T$ is a central inference problem in many scientific and engineering applications. 
For example, in vehicle navigation, one seeks to infer the vehicle trajectory $\bm{x}_1, \ldots, \bm{x}_T$ from noisy observations $\bm{y}_1, \ldots, \bm{y}_T$ such as signals obtained from GPS sensors~\cite{Zhao_Ochieng_Quddus_Noland_2003}.

\begin{figure}[tb]
  \includegraphics[width=\linewidth]{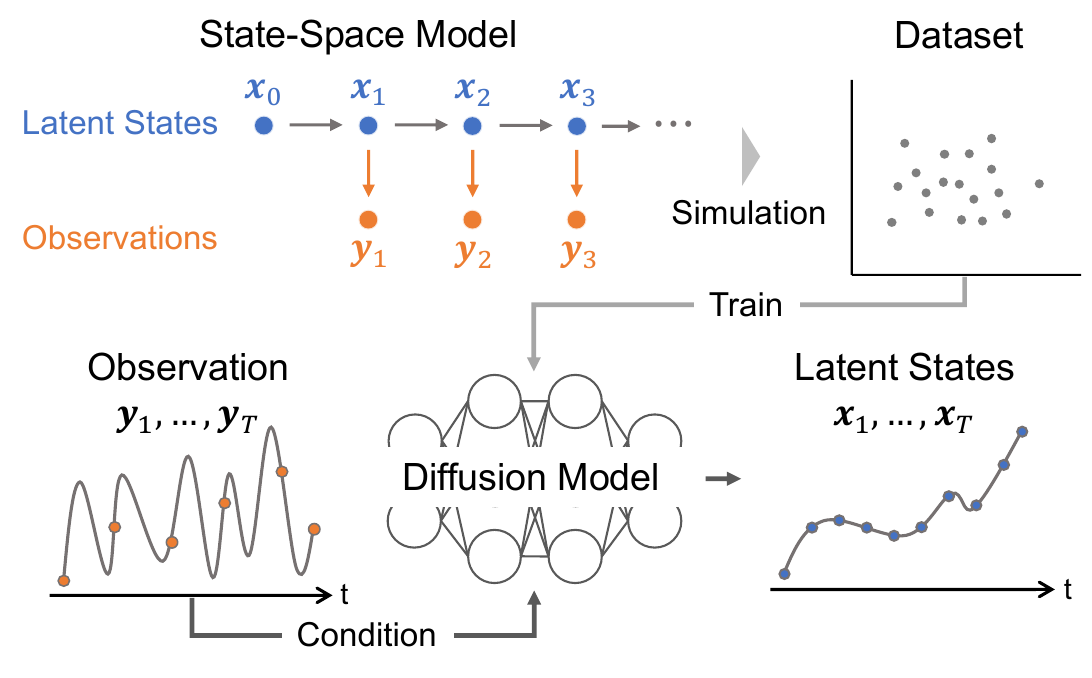}\\[-7mm]
  \caption{Overview of the training and inference procedure in Score-based Data Assimilation (SDA)~\cite{NEURIPS2023_SDA}. 
We first simulate samples from a given state-space model. 
These samples are used as a training dataset to train a diffusion model for sampling latent states $\bm{x}_1,..., \bm{x}_T$ conditioned on a given time-series observation $\bm{y}_1,..., \bm{y}_T$.}
  \label{abstract_}
  \vspace{-6mm}
\end{figure}

State-space models provide a fundamental framework for addressing this problem. 
This model consists of two components: 
(i) a latent state process that describes the temporal evolution of unobserved states $\bm{x}_t$ (illustrated by the transition chain $\cdots \rightarrow \bm{x}_{t-1} \rightarrow \bm{x}_t \rightarrow \bm{x}_{t+1} \rightarrow \cdots$ in the upper left of Fig.~\ref{abstract_}), and 
(ii) an observation process that maps each latent state $\bm{x}_t$ to an observable measurement $\bm{y}_t$ (illustrated by $\bm{x}_t \rightarrow \bm{y}_t$ in the same figure). 
By regarding a target system as the state-space model, one can infer the latent states $\bm{x}_1, \ldots, \bm{x}_T$ from noisy observations $\bm{y}_1, \ldots, \bm{y}_T$.

Nevertheless, the applicability of existing state-space modeling approaches remains limited. Classical (Extended and Ensemble) Kalman filters~\cite{Kalman_Filter, Extended_Kalman_Filter, Ensemble_Kalman_Filter} fundamentally rely on linear-Gaussian assumptions in  the latent state transition $\bm{x}_t \rightarrow \bm{x}_{t+1}$ and the observation model $\bm{x}_t \rightarrow \bm{y}_t$, which restricts their applicability to systems with complex dynamics. While particle filters~\cite{Particle_Filter} alleviate these modeling assumptions, their performance is known to degrade severely in high-dimensional systems.

To address this challenge, Rozet et al.~\cite{NEURIPS2023_SDA} proposed \emph{Score-based Data Assimilation} (SDA), a latent-state inference method for state-space models with high-dimensional and complex transition dynamics, leveraging diffusion models~\cite{DDIM}. Diffusion models are a powerful class of deep generative models demonstrating remarkable success in generating and sampling extremely high-dimensional data, such as images and videos.
SDA applies diffusion models to latent-state inference in state-space models. Specifically, the method trains a diffusion model on simulated data from the state-space model, and uses this model to sample latent states that are consistent with given observations (Fig.~\ref{abstract_}). Since this approach relies solely on simulated data from a given state-space model and does not require an explicit form of the latent-state transition, it is applicable to a broad class of state-space models in which assumptions such as linearity and Gaussianity do not hold. Moreover, by inheriting the ability of diffusion models to handle high-dimensional data, SDA naturally scales to high-dimensional state-space models.

However, as mentioned in the original paper (Section 5 in~\cite{NEURIPS2023_SDA}), SDA cannot be applied when the latent-state transition depends on unknown parameters. Many state-space models involve such unknown parameters in their latent-state dynamics. A representative example is a state-space model in which the latent state evolves according to a differential equation,
$
\frac{d\bm{x}(t)}{dt} = f\bigl(\bm{x}(t), \bm{\theta}\bigr). 
$
Such systems often involve unknown parameters $\bm{\theta}$ that govern the dynamics.
The original SDA formulation treats 
them as fixed and known and thus does not directly support joint inference of states and parameters.

This limitation is problematic in practical data assimilation, where 
unknown parameters often encode physical constants, operating conditions, and 
environment-specific properties. Reliable parameter estimates, together with latent states, enable simulator calibration and uncertainty quantification, improve forecasting by correcting transition dynamics, and provide interpretable quantities tied to domain knowledge~\cite{Smith, Ruiz, evensen2009data}. 

In this study, we propose a framework that enables SDA to handle latent states with unknown parameters. A key ingredient of the proposed framework is the incorporation of a self-organization method~\cite{Self_Organizing}, which has been primarily utilized in classical inference methods for state-space models, into SDA. By combining this classical technique with a modern SDA framework, it becomes possible to jointly infer latent states and parameters while maintaining the high training efficiency of SDA. We demonstrate the effectiveness of the proposed approach through 
the FitzHugh–Nagumo model, the Lorenz-63 system, and a high-dimensional Kolmogorov flow.

\section{Background}
\subsection{State-Space Modeling}
We formally describe the problem setting of state-space modeling.
Let
$
\bm{x}_{1:T} := (\bm{x}_1, \ldots, \bm{x}_T) \in \mathbb{R}^{d_{\mathcal{X}} \times T}
$
denote a trajectory of latent states evolving according to discrete-time stochastic dynamics,
$
\bm{x}_{t+1} \sim p(\bm{x}_{t+1} \mid \bm{x}_t).
$
In state-space modeling, observations are also given as a time series,
$
\bm{y}_{1:T} := (\bm{y}_1, \ldots, \bm{y}_T) \in \mathbb{R}^{d_{\mathcal{Y}} \times T},
$
assumed to be generated from a stochastic observation model,
$
\bm{y}_t \sim p(\bm{y}_t \mid \bm{x}_t).
$
The goal of state-space modeling is to sample a latent trajectory \( \bm{x}_{1:T} \) from the posterior distribution given observations $\bm{y}_{1:T}$. Specifically, we aim to obtain a sample from the posterior 
\begin{equation}\label{eq:posterior}
p(\bm{x}_{1:T} \mid \bm{y}_{1:T})
= \frac{\prod_{t=1}^T p(\bm{y}_t \mid \bm{x}_t)\, p(\bm{x}_{1:T})}{p(\bm{y}_{1:T})},
\end{equation}
which represents the distribution over latent state trajectories conditioned on the observed data.

The Kalman filter and particle filters are well-established methods in state-space modeling, enabling the sampling of latent trajectories from the posterior distribution \eqref{eq:posterior}. 
However, they are known to be difficult to scale to high-dimensional and nonlinear systems. 
The Kalman filter and its extensions (e.g., the extended and ensemble Kalman filters)~\cite{Kalman_Filter,Extended_Kalman_Filter,Ensemble_Kalman_Filter}
rely on linearization or Gaussian assumptions. 
Particle filters~\cite{Particle_Filter} provide a more general framework, 
but they often suffer from severe performance degradation in high-dimensional settings.

\subsection{Diffusion Models}

Diffusion models have emerged as a powerful tool for generating samples from target data distributions.
We review the technical background of diffusion models based on the formulation in~\cite{DDIM}.
 
\paragraph{Technical Background of Diffusion Models} 
Diffusion models consist of a forward diffusion process that gradually adds noise to data
and a reverse diffusion process that gradually removes noise in order to recover the data distribution.
In the forward diffusion process, noise is added to the target data $\bm{x} = \bm{x}^0$
from timestep $a = 0$ to $a = A$\footnote{Although it is common to use $t$ to denote time in diffusion processes,
we use $a$ to distinguish it from the time index of the state-space model.}
according to the stochastic differential equation
\begin{equation}\label{eq:forward_diffusion}
    d \bm{x}^a = f(a)\, \bm{x}^a \, da + g(a)\, dW,
\end{equation}
where $W$ is a standard Wiener process.
The drift and diffusion coefficients \( f(a) \) and \( g(a) \) are chosen such that
the SDE~\eqref{eq:forward_diffusion} transforms any initial distribution
into a normal distribution \( \mathcal{N}(\mu_A, I_A) \) at \( a = A \).
Sampling from the target distribution \( p(\bm{x}) = p(\bm{x}^0) \) can then be performed
by solving the corresponding reverse-time process~\cite{Anderson}
\begin{equation}\label{eq:backward_diffusion}
    d \bm{x}^a =
    \left\{
        f(a) - g^2(a)\, \nabla_{\bm{x}^a} \log p_a(\bm{x}^a)
    \right\} da
    + g(a)\, d\bar{W},
\end{equation}
backward in time from \( a = A \) to \( a = 0 \).
Here, \( \bar{W} \) denotes a backward Wiener process.
The reverse-time process requires the score function \( \nabla_{\bm{x}^a} \log p_a(\bm{x}^a) \),
whose analytical form is generally intractable.
To address this issue, we employ denoising score matching~\cite{Hyvarinen} and train a score network \( s_\phi(\bm{x}^a, a) \)
using a dataset \( \mathcal{D} \) consisting of samples drawn from the target distribution,
i.e., \( \mathcal{D} \sim p(\bm{x}) \).

\paragraph{Conditional Generation} Diffusion models can be readily extended to conditional generation.
Given a forward process~\eqref{eq:forward_diffusion} with an initial distribution $p(\bm{x} \mid \bm{y})$,
the corresponding reverse process~\eqref{eq:backward_diffusion}
involves the conditional score $\nabla_{\bm{x}^a} \log p(\bm{x}^a \mid \bm{y})$.
This conditional score can be estimated in essentially the same manner as described above.
During training, we require a dataset drawn from the joint distribution $p(\bm{x}, \bm{y})$,
i.e., $\mathcal{D} \sim p(\bm{x}, \bm{y})$.

In particular, when the explicit form of $p(\bm{y} \mid \bm{x})$ is specified,
conditional generation can be performed by training only the unconditional
score function $\nabla_{\bm{x}^a} \log p(\bm{x}^a)$.
This follows from the identity
\begin{equation*}
\nabla_{\bm{x}^a} \log p(\bm{x}^a \mid \bm{y})
=
\nabla_{\bm{x}^a} \log p(\bm{x}^a)
+
\nabla_{\bm{x}^a} \log p(\bm{y} \mid \bm{x}^a),
\end{equation*}
together with the fact that the second term can be evaluated using
 Tweedie's lemma~\cite{Chung2023} when $p(\bm{y} \mid \bm{x})$ is known.

\subsection{State-Space Modeling with Diffusion Models}\label{sda_}

Score-based Data Assimilation (SDA)~\cite{NEURIPS2023_SDA} is a method for latent state estimation~\eqref{eq:posterior}
that leverages recent advances in diffusion models.
In SDA, we first simulate a dataset
from a given state-space model (upper part of Fig.~\ref{abstract_}).
Using this dataset $\mathcal{D}$ as training data, we train a score function
$
\nabla_{\bm{x}^a_{1:T}} \log p(\bm{x}^a_{1:T} \mid \bm{y}_{1:T})
$
to run the reverse diffusion process~\eqref{eq:backward_diffusion}.
Finally, given an observation sequence $\bm{y}_1, \ldots, \bm{y}_T$,
we obtain samples from the posterior distribution~\eqref{eq:posterior}
using the trained score function or diffusion model (lower part of Fig.~\ref{abstract_}).
In SDA, we assume that the observation process $p(\bm{y}_{t} \mid \bm{x}_{t})$ follows a normal distribution
\begin{equation}\label{eq:observation_process}
  p(\bm{y}_t | \bm{x}_t) = \mathcal{N} ( \bm{y}_t | \mathcal{A}(\bm{x}_t), \Sigma),
\end{equation}
for a given observation operator $\mathcal{A}$, as is standard in data assimilation and state-space modeling~\cite{Kalman_Filter,Ensemble_Kalman_Filter}.
In this setting, we have access to an analytical form of the observation process. Therefore, for conditional generation $p(\bm{x}_{1:T} \mid \bm{y}_{1:T})$, it is sufficient to simulate the latent states $\bm{x}_{1:T}$ and use them as training data to train the unconditional score $\nabla_{\bm{x}^a_{1:T}} \log p(\bm{x}^a_{1:T})$ (Section II.B.b).

A key advantage of SDA is that it does not require assumptions on an explicit model form, as long as simulation from the given state-space model is possible. This is in contrast to the (Extended or Ensemble) Kalman filter, which imposes restrictive assumptions such as linearity and Gaussianity. Moreover, by inheriting strong sampling performance of diffusion models for high dimensional data, SDA is expected to scale well to high-dimensional state-space modeling. This stands in contrast to particle filters~\cite{Particle_Filter}, which are known to suffer from severe scalability issues in high-dimensional settings.

\begin{figure}[t]
  \centering
  \includegraphics[width=\linewidth]{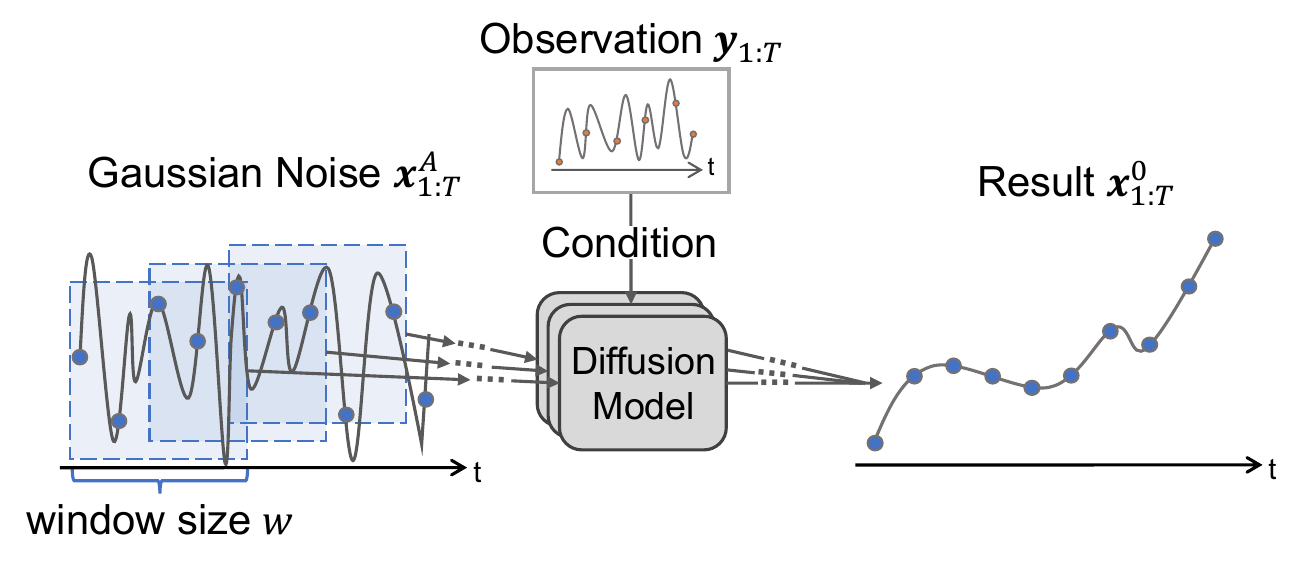}\\[-3mm]
  \caption{We train the diffusion model on trajectories that are shorter than 
  %those used during inference
  observations $\bm{y}_{1:T}$.
We refer to the training trajectory length as the \emph{window size}~$w$. The initial states $\bm{x}_{1:T}^A$ at $a = A$ of the reverse diffusion process are divided into short trajectories of length $w$, and each short trajectory is individually fed into the diffusion model. The final inferred latent state $\bm{x}_{1:T}^0$ is obtained by aggregating the outputs of these short trajectories.}
  \label{markov_blanket_inference}
  \vspace{-5mm}
\end{figure}

Another key property of SDA is its ability to reduce the simulation cost required
for training diffusion models by leveraging the Markov property of state-space
models. Figure~\ref{markov_blanket_inference} illustrates the intuition behind
this technique, while we refer the reader to~\cite{NEURIPS2023_SDA} for the
mathematical derivation. In a naïve application of SDA, especially when handling observations $\bm{y}_{1:T}$ with a long time-series length $T$, costly model simulations over long time horizons are required.
Rozet et al.~\cite{NEURIPS2023_SDA} mitigate this issue by exploiting the Markovian structure of the state-space model. Specifically, they show that the target score
$
\nabla_{\bm{x}^a_{1:T}} \log p(\bm{x}^a_{1:T})
$
can be factorized into local scores
$
\nabla_{\bm{x}^a_{i:i+w}} \log p(\bm{x}^a_{i:i+w}),
$
which can be estimated using model simulations over short segments $\bm{x}_{i:i+w}$ of length $w$, called the {\em window size}. This factorization enables efficient sampling for long time-series observations using only short segments of window size $w \ll T$.

However, the current SDA framework is not directly applicable when the state-space model contains unknown parameters. In many cases, state-space models---particularly the latent-state model $p(\bm{x}_{1:T})$---include parameters such as the initial conditions of the latent dynamics or the coefficients of the underlying differential equations. Jointly estimating both the state-space model parameters and the latent states is a fundamental problem in state-space modeling; however, this remains challenging within existing SDA frameworks. This difficulty in SDA is also discussed in Section~5 of~\cite{NEURIPS2023_SDA}.

\begin{figure}[t]
  \centering
  \includegraphics[width=\linewidth]{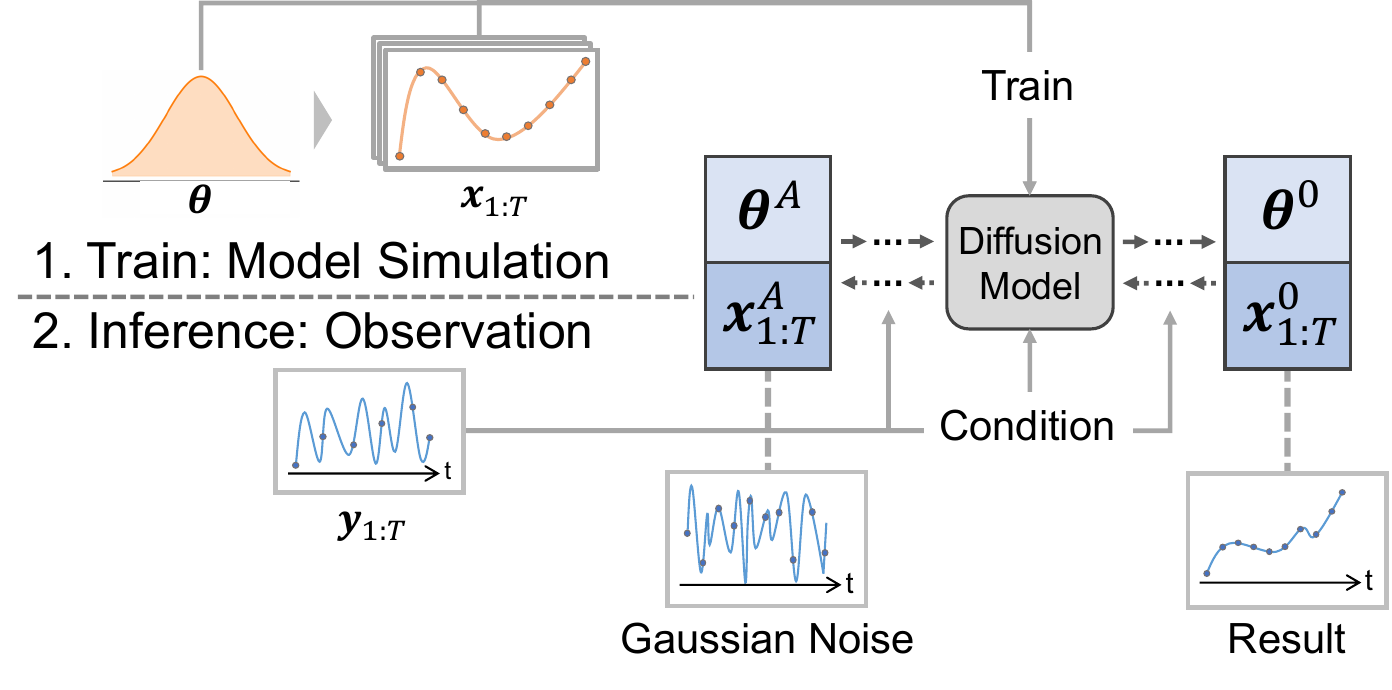}\\[-3mm]
  \caption{We simulate the latent states~$\bm{x}_{1:T}$ and the model parameter~$\theta$ to train the diffusion model. Subsequently, we use the trained model to jointly infer the latent states and model parameter conditioned on the observations~$\bm{y}_{1:T}$.}
  \label{joint_inference}
  \vspace{-5mm}
\end{figure}

\section{Self-Organizing Score-based Data Assimilation}

We propose a method for jointly estimating latent states and model parameters
in a state-space model within the framework of Score-based Data Assimilation (SDA).

\subsection{Problem Setting}
We consider a state-space model with a state transition density
$p(\bm{x}_{t+1} \mid \bm{x}_t, \bm{\theta})$, which depends on an unknown parameter $\bm{\theta}$
governing the system dynamics.
The objective of this paper is to infer the joint posterior distribution
of the latent states $\bm{x}_{1:T}$ and the parameter $\bm{\theta}$ given the observations
$\bm{y}_{1:T}$, namely, obtaining a sample from 
\begin{equation}\label{eq:goal_statespace}
p(\bm{x}_{1:T}, \bm{\theta} \mid \bm{y}_{1:T})
= \frac{
\prod_{t=1}^{T} p(\bm{y}_t \mid \bm{x}_t)\, p(\bm{x}_{1:T}, \bm{\theta})
}{
p(\bm{y}_{1:T})
}.
\end{equation}
For simplicity, throughout the study we assume the same Gaussian observation
process~\eqref{eq:observation_process} as in SDA.

\subsection{Joint Estimation of Latent States and Model Parameters via Model Simulation}\label{subsec:naive}

Building on the learning procedure in SDA, we train a diffusion model to infer latent states and model parameters~\eqref{eq:goal_statespace} using simulated data from the target state-space model.

Formally, we first simulate pairs $(\bm{x}_{1:T}, \bm{\theta})$ and use them as training data to train a target diffusion model (Fig.~\ref{joint_inference}, training). Here, noting that we assume a Gaussian observation process~\eqref{eq:observation_process}, it suffices to simulate pairs $(\bm{x}_{1:T}, \bm{\theta})$ consisting of the latent state and the model parameters for conditional generation.
At inference time, conditional generation given real observations is performed using the trained diffusion model, enabling us to infer latent-state trajectories and parameters that are consistent with the observations (Fig.~\ref{joint_inference}, inference).

\subsection{Handling Long Time-Series Observations}

However, the method described in Fig.~\ref{joint_inference} requires costly long time-series model simulations to cope with large-scale, long time-series observations.
The original SDA addresses this issue by exploiting the Markovian structure of the state-space model (Section II.C and Fig.~\ref{markov_blanket_inference}). This factorization relies on the assumption that the generation target $\bm{x}_{1:T}$ is Markovian. 
In the present setting, however, the generation target is $(\bm{x}_{1:T}, \bm{\theta})$, where $\bm{\theta}$ is shared across all time steps; therefore, the SDA technique cannot be applied directly.
To address this issue, we leverage a self-organization technique~\cite{Self_Organizing} developed in the context of classical state-space modeling.

\begin{figure}[t]
  \centering
  \includegraphics[width=1.0\linewidth]{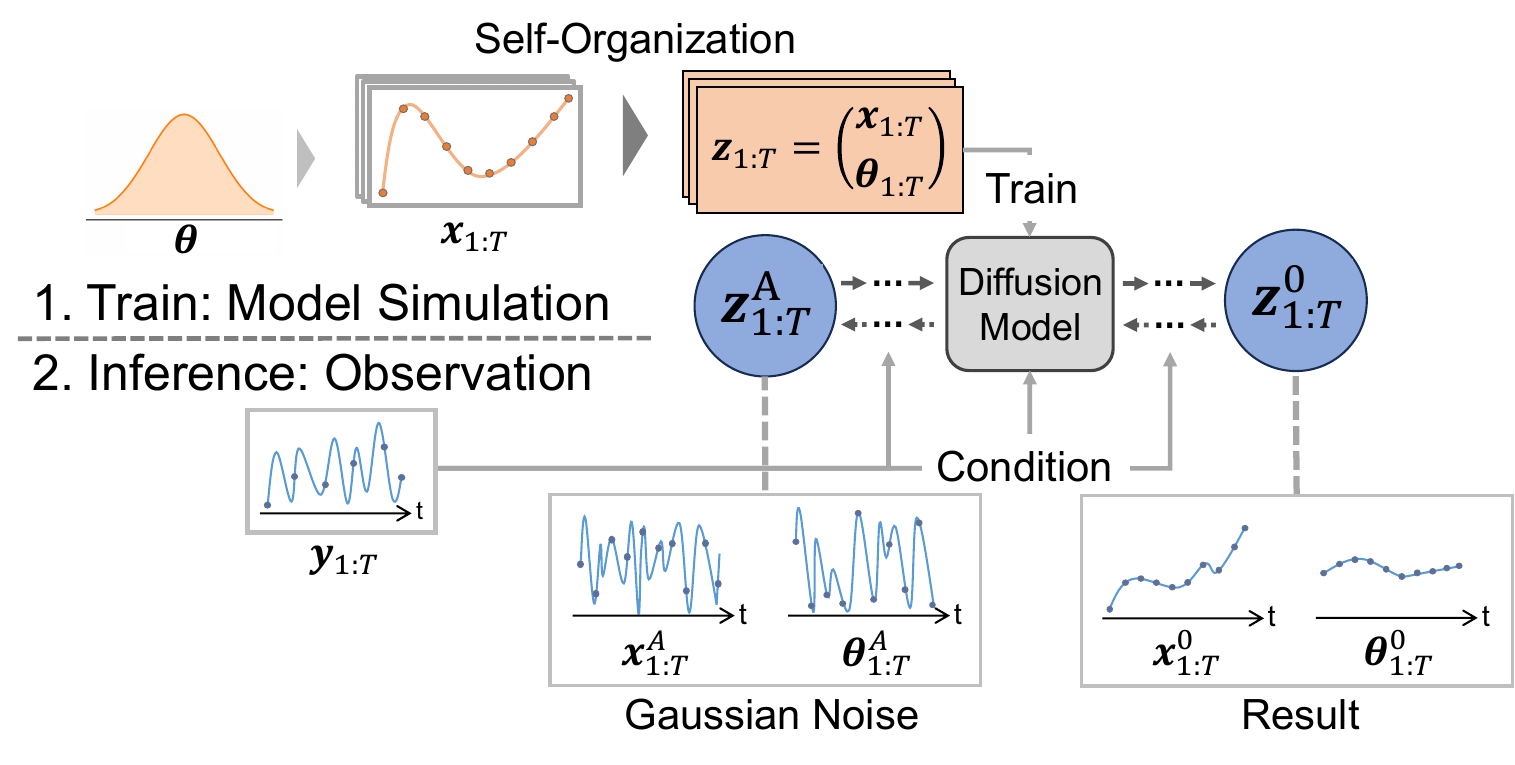}\\[-3mm]
  \caption{In contrast to the naive method described in Fig.~\ref{joint_inference}, we simulate the augmented latent states~$\bm{z}_{1:T}=(\bm{x}_{1:T},\bm{\theta}_{1:T})$ to train the diffusion model. As a result, the inferred model parameter is obtained as a trajectory~$\bm{\theta}_{1:T}$ rather than a static parameter $\bm{\theta}$.}
  \label{score_inference}
  \vspace{-5mm}
\end{figure}

\subsubsection{Self-Organization in State-Space Modeling}

Self-organization~\cite{Self_Organizing} is a classical approach, primarily used in particle filtering, for the joint estimation of the latent state $\bm{x}_t$ and an unknown parameter $\bm{\theta}$ in state-space models.
In this method, the model parameter $\bm{\theta}$ is extended to a time series $\bm{\theta}_{1:T}$ and embedded in the latent state $\bm{x_t}$ at each time step $t$:
\begin{equation}\label{eq:augmented}
\bm{z}_t =\begin{pmatrix}
\bm{x}_t \\
\bm{\theta}_t
\end{pmatrix}
\end{equation}

We treat $\bm{z}_t$ as a temporally evolving latent state, where $\bm{\theta}_t$ is either constant or follows a simple random walk, while $\bm{x}_t$ evolves according to the original latent dynamics $\bm{x}_t \rightarrow \bm{x}_{t+1}$:
\begin{align*}
    &\bm{x}_{t+1} \sim p(\bm{x}_{t+1} \mid \bm{x}_t, \bm{\theta}_t), \\
    \bm{\theta}_{t+1} &= \bm{\theta}_t \quad \bigl( + \bm{\epsilon}_t,\; \bm{\epsilon}_t \sim \mathcal{N}(\bm{0}, \bm{\Sigma}) \bigr).
\end{align*}

For the augmented latent state $\bm{z}_t$, we define the observation process $\bm{z}_t \rightarrow \bm{y}_t$ as
$
p(\bm{y}_t \mid \bm{z}_t) := p(\bm{y}_t \mid \bm{x}_t),
$
which again yields a state-space model. By applying a latent-state estimation method, such as particle filtering, to this augmented state-space model, one can estimate
the augmented latent state $\bm{z}_t = (\bm{x}_t, \bm{\theta}_t)^\top$, 
thereby obtaining both the model parameters and the latent states aligned with the observations $\bm{y}_{1:T}$.

\subsubsection{Incorporating Self-Organization into SDA}\label{subsec:SDAwithSelf}

By utilizing self-organization, we can extend the simulation cost reduction technique proposed in SDA to our setting of joint inference for the latent states and model parameters $(\bm{x}_{1:T}, \bm{\theta})$.

We first augmented latent space $\bm{x}_t$ to $\bm{z}_t := (\bm{x}_t, \bm{\theta}_t)^\mathsf{T}$ in the manner of self-organization described in the last section \eqref{eq:augmented}. We next collect a large number of simulated datasets $\bm{z}_{1:T} =( \bm{x}_t, \bm{\theta}_t)^\mathsf{T}$ 
(Fig.~\ref{score_inference}, training). Using these as training data, we train a diffusion model that generates the augmented latent-state trajectory $\bm{z}_{1:T}$ conditioned on observations, achieving joint inference of parameters and latent states (Fig.~\ref{score_inference}, inference). As the model parameters are finally obtained in the form of a time series $\bm{\theta}_{1:T}$, summary statistics of this sequence (e.g., the mean) are used as the final parameter estimates.

Unlike the naive method described in Fig.~\ref{joint_inference}, the approach shown in Fig.~\ref{score_inference} considers a Markovian time series, $ \cdots \rightarrow \bm{z}_{t-1} \rightarrow \bm{z}_t \rightarrow \bm{z}_{t+1} \rightarrow \cdots$, as the generation target. Therefore, we can directly apply the simulation-budget reduction techniques in SDA, relying on the Markov property of the generation target. This enables joint inference of the latent state $\bm{x}_t$ and the parameter $\bm{\theta}$ for large-scale time-series observations with fewer simulation budgets, which is the desired outcome. We will examine in the next section how short a simulation length is actually sufficient.

\section{Experiments}

We evaluate the proposed method on the FitzHugh--Nagumo ~\cite{FITZHUGH, Nagumo} and the Lorenz-63 system~\cite{Lorenz63}, which are ordinary differential equation models in neuroscience and atmospheric science, respectively. Finally, we demonstrate its scalability using a high-dimensional Kolmogorov flow~\cite{Kolmogorov_Flow}.

\subsection{FitzHugh-Nagumo Model and Lorenz-63 System}\label{experiments:both_systems}
We first demonstrate the proposed approach on two ODE models: FitzHugh--Nagumo (FHN) model
\begin{equation}
\begin{aligned}
\dot{u} &= u - \frac{u^3}{3} - v + I, \\
\tau\,\dot{v} &= u + a - b v
\end{aligned}
\label{eq:FHN}
\end{equation}
and Lorenz--63 defined as
\begin{equation}
\begin{aligned}
\dot{x} &= \sigma (y - x), \\
\dot{y} &= x(\rho - z) - y, \\
\dot{z} &= xy - \beta z.
\end{aligned}
\label{eq:Lorenz63}
\end{equation}

We consider the dynamics
$
\bm{x}_t = (u_t, v_t)^\top \in \mathbb{R}^2
$
and
$
\bm{x}_t = (x_t, y_t, z_t)^\top \in \mathbb{R}^3,
$
obtained by numerically solving the ODEs \eqref{eq:FHN} and
\eqref{eq:Lorenz63} with step sizes of $0.4$ (FHN) and $0.05$ (Lorenz--63),
respectively.
By treating them as latent states, 
we perform inference on both the latent states and the unknown parameters that govern these equations.
While ODEs~\eqref{eq:FHN} and \eqref{eq:Lorenz63} include three unknown parameters,
$\bm{\theta} = (a, b, \tau)$ and $\bm{\theta} = (\rho, \sigma, \beta)$, respectively,
for simplicity we consider a single unknown model parameter and fix the remaining two.
Specifically, for FHN model, we fix $b = 0.2$ and $\tau = 1.0$ and infer $a$.
For the Lorenz--63 system, we fix $\sigma = 10$ and $\beta = 8/3$ and infer $\rho$.  
In the observation process, only the first component ($u_t$ for FHN and $x_t$ for Lorenz--63) is observed every 8 time steps, corrupted by Gaussian noise $\mathcal{N}(0, 0.05^2)$.

\begin{figure}[t]
  \centering
  \includegraphics[width=\linewidth]{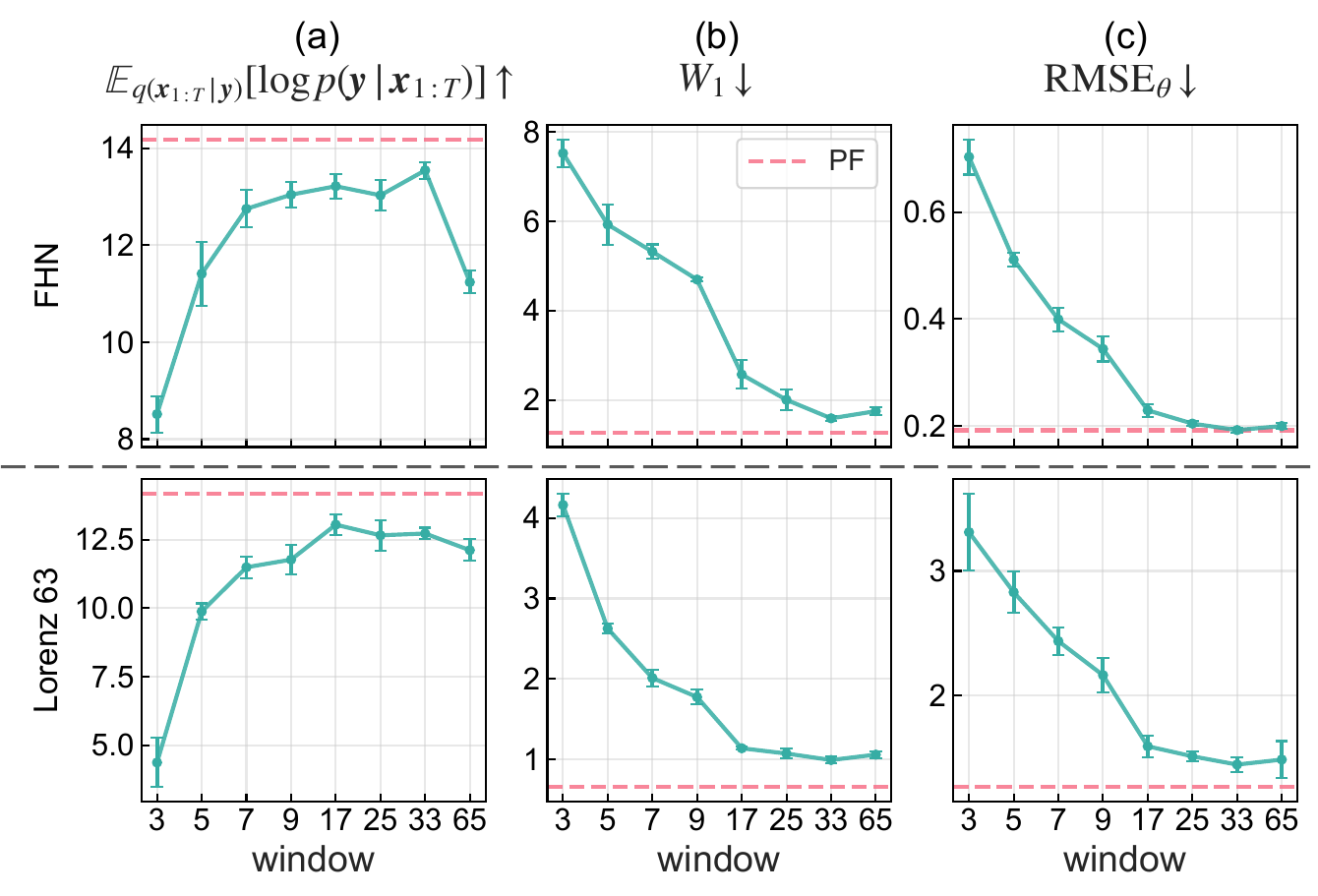}\\[-3mm]
  \caption{Evaluation metrics across window sizes $w \in \{3, 5, 7, 9, 17, 25, 33, 65\}$, compared with the particle filter (PF; red dashed line), for FHN (top row) and Lorenz-63 (bottom row). Panels show: (a) expected log-likelihood, (b) Wasserstein distance for the latent states, and (c) RMSE of the model parameter $\theta$.}
  \label{lorenz63:quantitative}
  \vspace{-5mm}
\end{figure}
\begin{figure}[t]
  \centering
  \includegraphics[width=\linewidth]{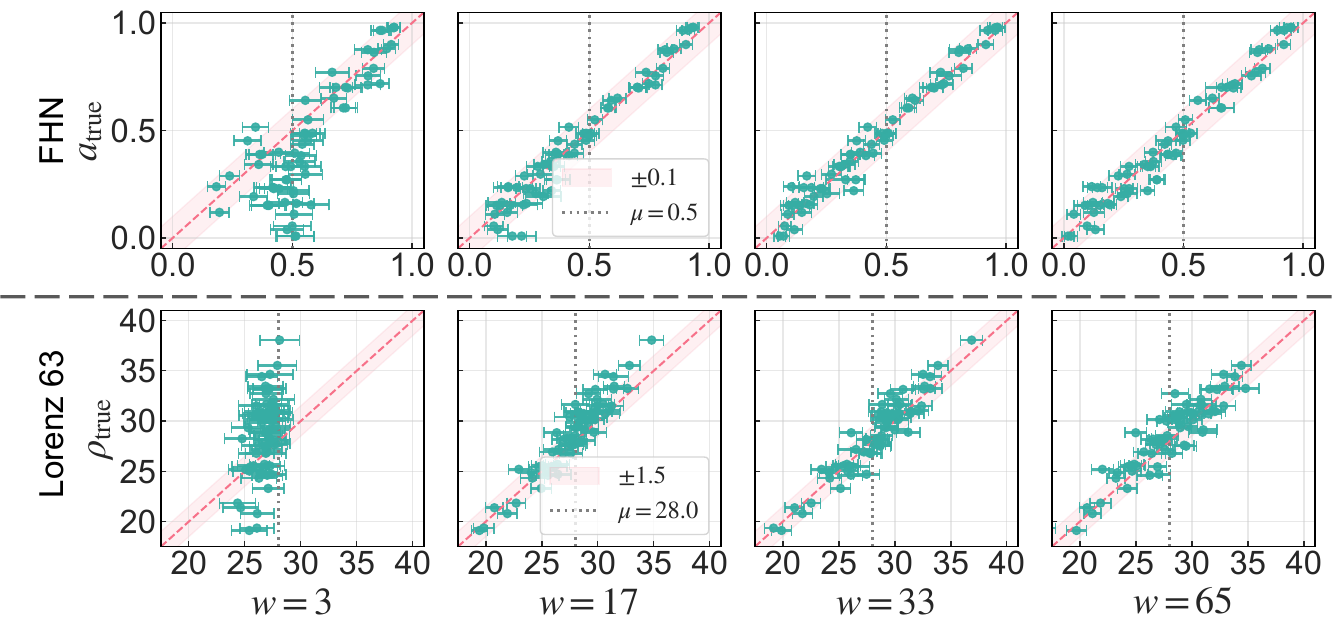}\\[-3mm]
    \caption{Comparison between the ground-truth and inferred model parameters across window sizes $w \in \{3, 17, 33, 65\}$, for FHN (top row) and Lorenz-63 (bottom row). The red dashed line represents the ideal estimation ($y=x$), and the vertical dotted line marks the mean of the training distribution ($\mu=0.5$ for FHN and $\mu=28.0$ for Lorenz-63).}
  \label{lorenz63:rho_scatter_by_window}
  \vspace{-4mm}
\end{figure}

We generate a dataset of $4096$ pairs $(\bm{x}_{1:T},\theta)$ by sampling model parameters from the prior ($a\sim \mathrm{Uniform}(0,1.0)$ for FHN and $\rho\sim\mathcal{N}(28.0,4.0^2)$ for Lorenz-63) and simulating the system for $1024$ steps. We split the dataset into 80\% training, 10\% validation, and 10\% test sets. 
In our method, we train on shorter trajectories than those used during inference; we refer to the training-trajectory length as the window size~$w$ (Fig.~\ref{markov_blanket_inference}).
We evaluate our method using $64$ observations generated by applying the observation process to $T=65$ trajectory segments extracted from test-set trajectories. 

Since the latent states of both systems are low-dimensional, we obtain a reliable reference posterior $q(\bm{x}_{1:T},\bm{\theta}_{1:T}\mid \bm{y}_{1:T})$ using a particle filter (PF)~\cite{Particle_Filter}. We then quantitatively evaluate our method by comparing the inferred posterior $p(\bm{x}_{1:T},\bm{\theta}_{1:T}\mid \bm{y}_{1:T})$ with the reference posterior $q(\bm{x}_{1:T},\bm{\theta}_{1:T}\mid \bm{y}_{1:T})$. We use three metrics: a) the expected log-likelihood $\mathbb{E}_{q(\bm{x}_{1:T}\mid \bm{y}_{1:T})}[\log p(\bm{y}_{1:T} \mid \bm{x}_{1:T})]$; b) the Wasserstein distance $W_1(p,q)$; c) the root mean squared error ($\mathrm{RMSE}$). For the expected log-likelihood, larger is better, whereas for $W_1(p,q)$ and $\mathrm{RMSE}$, lower is better. 
We also evaluate the performance of model parameter estimation by visualizing the true and inferred model parameters in two dimensions.

Fig.~\ref{lorenz63:quantitative} presents the quantitative evaluation of the inferred samples. To show the best performance for each evaluation metric, we also report the results of comparisons between samples drawn from the reference posterior obtained by the particle filter (PF in Fig.~\ref{lorenz63:quantitative}). Across all metrics, our method achieves performance close to  
the best performance (PF)
for FHN when $w\ge25$ and for Lorenz-63 when $w\ge17$.
Some results indicate a degradation in performance at $w=65$, especially for the expected loss in the FHN model; this is likely because increasing the window size makes the problem higher-dimensional and therefore harder to estimate accurately.
Fig.~\ref{lorenz63:rho_scatter_by_window} visualizes the inferred model parameters and compares them with the corresponding true parameters.
For small window sizes (e.g., $w=3$), the inferred parameters cluster around the mean parameters used for training ($a=0.5$ for FHN, $\rho=28.0$ for Lorenz-63); this can be explained by the fact that, with a small window size, sufficient information to infer the parameters could not be obtained from the observations, causing the inferred parameters to be distributed like a prior. Whereas for $w\ge17$, they concentrate near the ground-truth values. Overall, these results indicate that the proposed framework can correctly infer both latent states and parameters
using window sizes $w = 25$ (FHN) or $w = 17$ (Lorenz-63).
\subsection{Kolmogorov Flow}

The Kolmogorov flow is governed by the incompressible Navier--Stokes equations:
\begin{equation}
        \dot{\mathbf{u}} = -\mathbf{u}\nabla\mathbf{u}+\frac{1}{Re}\nabla^2\mathbf{u}-\frac{1}{\rho_{\mathrm{kol}}}\nabla p+\mathbf{f}, ~~0=\nabla\cdot\mathbf{u},
\end{equation}
where $Re$ is the Reynolds number, $\rho_{\mathrm{kol}}$ the fluid density,
$p$ the pressure field, and $\mathbf{f}$ an external forcing.
The latent state $\bm{x}_t$ is defined as a snapshot of the two-dimensional
velocity field $\mathbf{u}$ on a $256\times256$ grid, solved by \texttt{jax-cfd} library~\cite{Jax_CFD}. 
We jointly infer $\bm{x}_t$ and the model parameter $\theta = Re$, while all
other parameters are fixed.

We generate a training data of $8192$ pairs $(\bm{x}_{1:T},Re)$ where $Re\sim \mathrm{Uniform}(10,200)$.
The simulated fields are then downsampled to $64\times64$ to form the latent states. Each trajectory has length of $64$ steps with time step $\Delta=0.2$. We split the dataset into 80\% training, 10\% validation, and 10\% test sets. 
We assume that observations are obtained by downsampling the latent state from $64\times64$ to $8\times8$ and adding Gaussian noise $\mathcal{N}(0, 0.1^2)$.
For evaluation, we extract segments of length $T=33$ from the test-set trajectories and generate observations by applying the same observation process.

\begin{figure}[t]
  \centering
  \includegraphics[width=\linewidth]{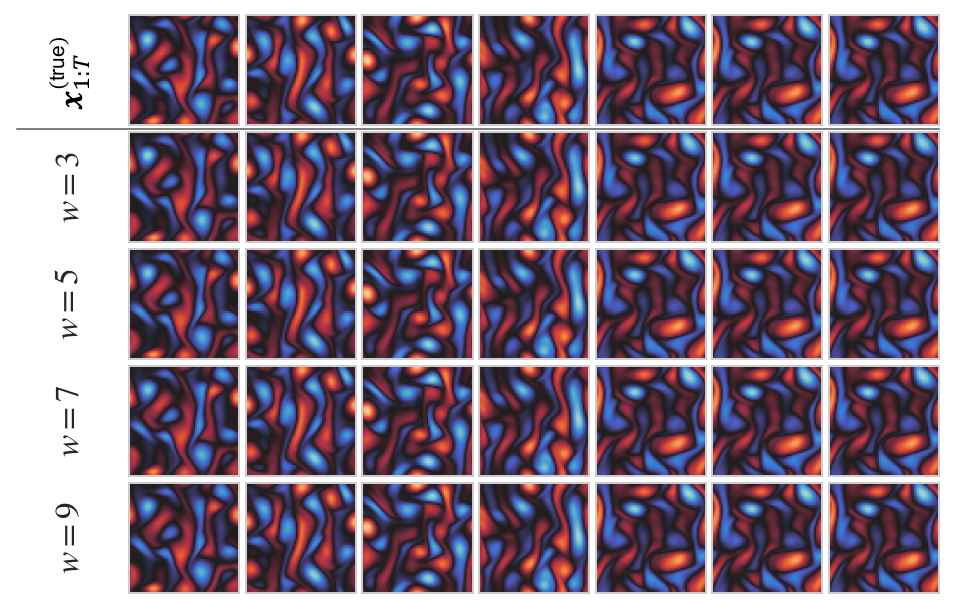}\\[-3mm]
  \caption{
  Visualization of the inferred latent states $\bm{x}_{1:T}$ for window sizes $w\in\{3, 5, 7, 9\}$, with the ground-truth $\bm{x}_{1:T}^{(\text{true})}$ (top row). Across all window sizes, our method recovers the ground-truth trajectory.}
  \label{kolmogorov_x_qualitative}
\end{figure}

\begin{figure}[t]
  \centering
  \includegraphics[width=\linewidth]{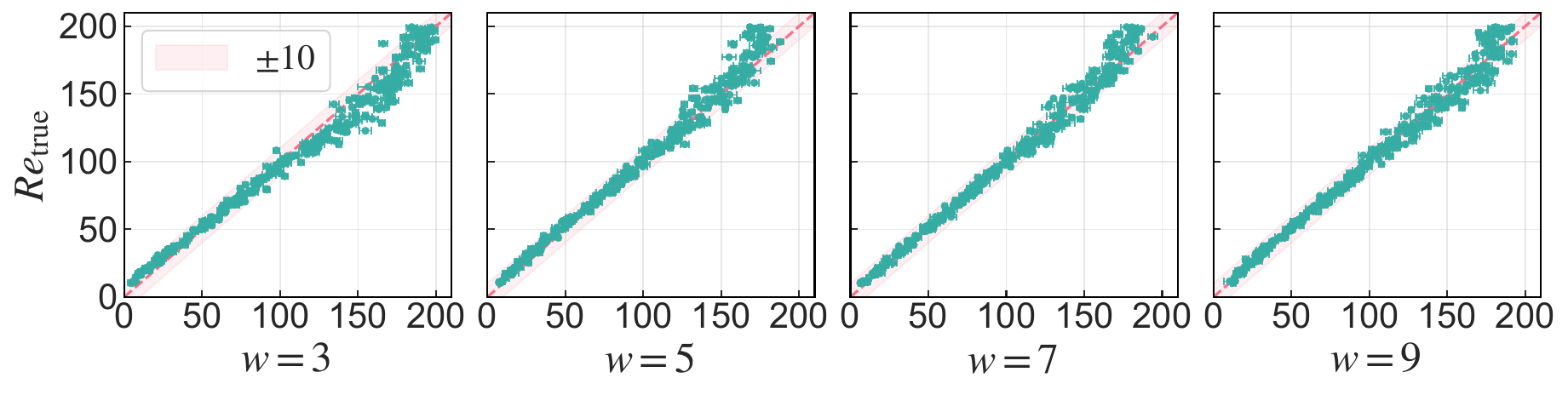}\\[-3mm]
  \caption{Comparison of the ground-truth Reynolds number and the inferred Reynolds number across window sizes $w \in \{3, 5, 7, 9\}$. The dashed line represents the identity line ($y=x$).}
  \label{kolmogorov_reynolds_scatter_by_window}
  \vspace{-2mm}
\end{figure}

Fig.~\ref{kolmogorov_x_qualitative} shows the inferred latent states for observations generated from the ground-truth trajectory $\bm{x}_{1:T}^{(\mathrm{true})}$. This qualitative result indicates that our method can recover the ground-truth trajectory. 
Fig.~\ref{kolmogorov_reynolds_scatter_by_window} shows the inferred Reynolds numbers 
compared with the true parameters. 
The results qualitatively indicate that the proposed method successfully recovers the true parameters.
This confirms that the proposed method can successfully infer the true parameters.
\begin{table}[ht]
    \renewcommand{\arraystretch}{1.3}
    \caption{Comparison of performance metrics across window sizes $w$.
    Values are reported as Means
    (STDs)
    over $900$ runs. }
    \begin{center}
        \begin{tabular}{c c c c}
            \hline
            $w$ & $\mathrm{RMSE}_{x}\downarrow$ & $\mathrm{RMSE}_{Re}\downarrow$ & $D_{\mathrm{equation}}\downarrow$ \\
            \hline
            3 & 0.056\,{\color{gray}(0.001)} & 0.144\,{\color{gray}(0.004)} & \textbf{4.05}\,{\color{gray}(0.09)} \\
            5 & \textbf{0.052}\,{\color{gray}(0.001)} & \textbf{0.107}\,{\color{gray}(0.003)} & 4.16\,{\color{gray}(0.10)} \\
            7 & 0.054\,{\color{gray}(0.001)} & 0.121\,{\color{gray}(0.003)} & 4.45\,{\color{gray}(0.11)} \\
            9 & 0.060\,{\color{gray}(0.001)} & 0.134\,{\color{gray}(0.004)} & 4.78\,{\color{gray}(0.12)} \\
            \hline
        \end{tabular}
        \label{tab:window_metrics_best}
    \end{center}
    \vspace{-5mm}
\end{table}

For quantitative evaluation, we measure the root mean squared error ($\mathrm{RMSE}$) between the ground truth and the inferred samples for both the latent states ($\mathrm{RMSE}_{x}$) and the Reynolds number ($\mathrm{RMSE}_{Re}$). We also assess dynamical consistency using the Equation Loss $D_{\mathrm{equation}}$~\cite{Equation_Loss}. For all metrics, lower values indicate better performance. 
Table~\ref{tab:window_metrics_best} summarizes the results across window sizes. 
We can confirm that smaller windows yield better accuracy. In particular, 
$w=5$ achieves the best overall performance, attaining the lowest $\mathrm{RMSE}_{x}$ and $\mathrm{RMSE}_{Re}$ while keeping $D_{\mathrm{equation}}$ competitive (the minimum $D_{\mathrm{equation}}$ is achieved at $w=3$).
This tendency may stem from the overwhelmingly high dimensionality of the Kolmogorov flow.
Increasing the window size $w$ necessitates
a greater amount of simulation data (under the full window setting $w = T$, the data dimension is
$64 \times 64 \times 33 = 135{,}168$); consequently, under a fixed simulation budget, a larger window size may result in degraded performance.
In contrast, a smaller window size partitions the data into finer segments, thereby increasing the number of available training samples and leading to improved inference performance.

\section{Conclusion}

We introduced Self-Organizing Score-based Data Assimilation, a method for jointly inferring latent states and parameters for general state-space models, scaling to high-dimensional systems while avoiding strong distributional assumptions. 
Key ingredient of the proposed method 
is an incorporation of self-organization technique---originally developed in the context of the particle filter---into the diffusion model, enabling inference on long observation trajectories with high training efficiency. We demonstrated the effectiveness of our approach 
on three dynamical systems, the FitzHugh--Nagumo, Lorenz-63 systems, and the Kolmogorov flow with the data dimension on the order of several hundred thousand.

\section*{Acknowledgment}

We used AI tools only for English proofreading; we are responsible for all scientific content and the manuscript. This work was supported by ACT-X-JPMJAX25C, JSPS KAKENHI JP22H05180, JP24K20750 and JP25H01454

\bibliographystyle{IEEEtran}
\bibliography{refs}

@inproceedings{NEURIPS2023_SDA,
    author = {Rozet, Fran\c{c}ois and Louppe, Gilles},
    booktitle = {Advances in Neural Information Processing Systems},
    
    title = {Score-based Data Assimilation},
    volume = {36},
    pages = {40521--40541},
year = {2023}
}

@article{Zhao_Ochieng_Quddus_Noland_2003,
    title={An Extended Kalman Filter Algorithm for Integrating GPS and Low Cost Dead Reckoning System Data for Vehicle Performance and Emissions Monitoring},
    volume={56},
    journal={Journal of Navigation},
    author={Zhao, L. and Ochieng, W. Y. and Quddus, M. A. and Noland, R. B.},
    year={2003},
    pages={257–275}
}

@article{Kalman_Filter,
    author = {Kalman, R. E.},
    title = {A New Approach to Linear Filtering and Prediction Problems},
    journal = {Journal of Basic Engineering},
    volume = {82},
    pages = {35-45},
    year = {1960},
}

@book{Extended_Kalman_Filter,
    title={Application of statistical filter theory to the optimal estimation of position and velocity on board a circumlunar vehicle},
    author={Smith, Gerald L and Schmidt, Stanley F and McGee, Leonard A},
    volume={135},
    year={1962},
}

@article{Ensemble_Kalman_Filter,
    title={Sequential data assimilation with a nonlinear quasi-geostrophic model using Monte Carlo methods to forecast error statistics},
    author={Evensen, Geir},
    journal={Journal of Geophysical Research: Oceans},
    volume={99},
    pages={10143--10162},
    year={1994},
}

@article{Particle_Filter,
    author = {N.J. Gordon  and D.J. Salmond  and A.F.M. Smith },
    title = {Novel approach to nonlinear/non-Gaussian Bayesian state estimation},
    journal = {IEE Proceedings F (Radar and Signal Processing)},
    volume = {140},
    pages = {107-113},
    year = {1993},
}

@inproceedings{DDIM,
    title={Denoising Diffusion Implicit Models},
    author={Jiaming Song and Chenlin Meng and Stefano Ermon},
    booktitle={International Conference on Learning Representations},
    year={2021},
}

@article{Self_Organizing,
    title={A self-organizing state-space model},
    author={Kitagawa, Genshiro},
    journal={Journal of the American Statistical Association},
    pages={1203--1215},
    year={1998},
}

@article {Lorenz63,
    author={Edward N. Lorenz},
    title={Deterministic Nonperiodic Flow},
    journal={Journal of Atmospheric Sciences},
    year={1963},
    volume={20},
    pages={130 - 141},
}

@article{Kolmogorov_Flow,
    title={Invariant recurrent solutions embedded in a turbulent two-dimensional Kolmogorov flow},
    volume={722},
    journal={Journal of Fluid Mechanics},
    author={Chandler, Gary J. and Kerswell, Rich R.},
    year={2013},
    pages={554–595}
}

@article{Anderson,
    title = {Reverse-time diffusion equation models},
    journal = {Stochastic Processes and their Applications},
    volume = {12},
    pages = {313-326},
    year = {1982},
    author = {Brian D.O. Anderson},
}

@article{Hyvarinen,
  title={Estimation of non-normalized statistical models by score matching.},
  author={Hyv{\"a}rinen, Aapo and Dayan, Peter},
  journal={Journal of Machine Learning Research},
  volume={6},
  year={2005}
}

@inproceedings{Chung2023,
    title={Diffusion Posterior Sampling for General Noisy Inverse Problems},
    author={Hyungjin Chung and Jeongsol Kim and Michael Thompson Mccann and Marc Louis Klasky and Jong Chul Ye},
    booktitle={International Conference on Learning Representations },
    year={2023},
}

@article{Jax_CFD,
    author = {Dmitrii Kochkov  and Jamie A. Smith  and Ayya Alieva  and Qing Wang  and Michael P. Brenner  and Stephan Hoyer },
    title = {Machine learning–accelerated computational fluid dynamics},
    journal = {Proceedings of the National Academy of Sciences},
    volume = {118},
    pages = {e2101784118},
year = {2021}
}

@article{Equation_Loss,
    title = {A physics-informed diffusion model for high-fidelity flow field reconstruction},
    journal = {Journal of Computational Physics},
    volume = {478},
    pages = {111972},
    year = {2023},
    author = {Dule Shu and Zijie Li and Amir {Barati Farimani}},
}

@article{FITZHUGH,
title = {Impulses and Physiological States in Theoretical Models of Nerve Membrane},
journal = {Biophysical Journal},
volume = {1},
pages = {445-466},
year = {1961},
author = {Richard FitzHugh}
}

@ARTICLE{Nagumo,
  author={Nagumo, J. and Arimoto, S. and Yoshizawa, S.},
  journal={Proceedings of the IRE}, 
  title={An Active Pulse Transmission Line Simulating Nerve Axon}, 
  year={1962},
  volume={50},
  pages={2061-2070}
}

@article{Smith,
    author = {Smith, P. J. and Thornhill, G. D. and Dance, S. L. and Lawless, A. S. and Mason, D. C. and Nichols, N. K.},
    title = {Data assimilation for state and parameter estimation: application to morphodynamic modelling},
    journal = {Quarterly Journal of the Royal Meteorological Society},
    volume = {139},
    pages = {314-327},
    year = {2013}
}

@article{Ruiz,
  title={Estimating Model Parameters with Ensemble-Based Data Assimilation: A Review},
  author = {Ruiz, Juan Jose and Pulido, Manuel and Miyoshi, Takemasa},
  journal={Journal of the Meteorological Society of Japan},
  volume={91},
  pages={79-99},
  year={2013},
}

@book{evensen2009data,
  title={Data assimilation: the ensemble Kalman filter},
  author={Evensen, Geir},
  year={2009},
  publisher={Springer}
}

\end{document}